\newcommand{\beq}{\begin{equation}}
\newcommand{\eeq}{\end{equation}}
\newcommand{\beqa}{\begin{eqnarray}}
\newcommand{\eeqa}{\end{eqnarray}}

\newcommand{\M}{{\cal M}}
\newcommand{\CM}{C_{\cal M}}
\newcommand{\CMtwo}{C_{\cal M}^2}
\newcommand{\K}{{\cal K}}
\newcommand{\CKtwo}{C_{\cal K}^2}
\newcommand{\Ctwo}[1]{C_{#1}^2}
\newcommand{\Pm}{P_{\cal  M}(m)}
\newcommand{\Pmtwo}{P_{\cal  M}^2(m)}
\newcommand{\levels}{d}
\newcommand{\particles}{N}
\newcommand{\Dim}{M}
\newcommand{\ketstate}{|\psi \rangle}
\newcommand{\opx}{\hat{\sigma}_x}
\newcommand{\opy}{\hat{\sigma}_y}
\newcommand{\opz}{\hat{\sigma}_z}

\documentclass[pra,superscriptaddress,showpacs,twocolumn]{revtex4}

\usepackage{graphicx}

\begin{document}

\title{Certainty relations between local and nonlocal observables}

\author{R. Garcia Diaz}
\affiliation{School of Information and Communication Technology,
Royal Institute of Technology (KTH), Electrum 229, SE-164 40 Kista,
Sweden}

\author{J. L. Romero} \affiliation{School of Information and Communication Technology,
Royal Institute of Technology (KTH), Electrum 229, SE-164 40 Kista,
Sweden}

\author{G. Bj\"{o}rk}
\affiliation{School of Information and Communication Technology,
Royal Institute of Technology (KTH), Electrum 229, SE-164 40 Kista,
Sweden}

\author{M. Bourennane}
\affiliation{Fysikum, Stockholm University, SE-106 91 Stockholm,
Sweden}

\date{\today}

\begin{abstract}
We demonstrate that for an arbitrary number of identical particles,
each defined on a Hilbert-space of arbitrary dimension, there exists
a whole ladder of relations of complementarity between local, and
every conceivable kind of joint (or nonlocal) measurements. E.g.,
the more accurate we can know (by a measurement) some joint property
of three qubits (projecting the state onto a tripartite entangled
state), the less accurate some other property, local to the three
qubits, become. We also show that the corresponding complementarity
relations are particularly tight for particles defined on prime
dimensional Hilbert spaces.
\end{abstract}


\maketitle

\section{Introduction}

Uncertainty and complementarity are two intriguing and fascinating
concepts inherent to quantum theory. Both phenomena can be traced to
the linearity of the Schr\"{o}dinger equation. Undoubtedly, the most
well-known relation originating from these concepts is Heisenberg's
uncertainty relation \cite{Heisenberg}. Traditionally, it is
presented for canonical observables, such as position and momentum
of a particle. Observables defined in a finite Hilbert space are not
canonical \cite{Wintner,Wielandt} and obey a related uncertainty
relation, the Schr\"{o}dinger-Robertson (S-R) uncertainty relation
\cite{Schrodinger}. Unfortunately the latter leads to a trivial
lower bound for the uncertainty product, namely zero for any
eigenstate of either of the observables. The nontrivial class of
states associated with a S-R uncertainty relation are the
intelligent states \cite{Kolodziejczyk}. These have the property
that they fulfill the S-R uncertainty relation with equality.
However, recently it was pointed out that these states do not give
the smallest uncertainty product for any given uncertainty
(variance) of one of the observables \cite{Pegg}. Therefore, it
seems like the mathematically well defined intelligent states have
little physical significance.

Other kinds of complementarity relations have also been discussed
\cite{Bialynicki,Deutsch,Uffink,Krauss,Maassen,Bialynicki 2}, some
more general, but many focusing on the wave-particle duality of
quantum mechanics \cite{Wootters
1,Bartell,Greenberger,Mandel,Jaeger,Englert 1,Bjork,Bjork 2,Englert
2}. In particular for two-state systems, elegant relations between
two complementary observables have been derived
\cite{Greenberger,Mandel,Jaeger,Englert 1} (often stated in terms of
which-path predictability or distinguishability and interference
visibility). Recently, generalizations of these relations have been
made to larger Hilbert spaces \cite{Durr,Kaszlikowski} (and the
corresponding relations have then usually been discussed in terms of
distinguishability and multi-path ``interference-visibility'').
Another line of generalizations of complementarity relations has
targeted ``simultaneous'' measurements of multiple observables
\cite{Ivanovic,Wootters 2,Larsen,Luis}. Specifically, Zeilinger has
proposed that, in fact, pure states uniquely correspond to
measurable discrete (e.g., binary) propositions \cite{Zeilinger}.
Together with Brukner, he has also derived an information invariant,
that unfortunately only holds for Hilbert spaces of a prime, or
prime to an integer power, dimension \cite{Brukner}. This invariant
is based on the measurement outcome of a set of mutually unbiased
Hermitian operators, as defined by Ivanovi\'{c} \cite{Ivanovic} and
extended by Wootters and Fields \cite{Wootters 2}.

In the context of complementarity, it has been noticed that there
also exists a complementarity relation between observation of
single-particle interference and two-particle interference
\cite{Kaszlikowski,Jaeger 2} (requiring a minimal Hilbert-space
dimension of four). In this paper we extend this idea, combine it
with the aforementioned multiobservable complementarity relations,
and show that there exists, in fact, a whole set of
complementarity relations between single-particle, two-particle,
three-particle, etc. interference (observed through a proper
observable), both if the observations are taken pairwise, or
all-together. Hence, the complementarity between single- and
two-photon interference is just a first step of a ladder of
similar relations.

\section{Preliminaries}

In this work we shall consider $\particles$ identical particles (or
subsystems), each defined on a $\levels$-dimensional Hilbert space
${\cal H}_\levels$. The Hilbert space of the composite system (all
the particles) is hence of dimension $\Dim = \levels^\particles$. In
this $\Dim$-dimensional space we call two observables $\K$ and $\M$
mutually unbiased (MUB) if all their respective, complete,
orthonormal eigenvectors fulfil \beq
|\langle{\K,k}|{\M,m}\rangle|^2=\frac{1}{\Dim} \ \forall \ k,m = 1,
\ldots , \Dim  . \eeq We already know that the set of mutually
unbiased bases (MUB) can be at most $\Dim+1$. It is also known that
if $\levels$ is prime (or $\Dim$ is a prime), the maximal number of
MUB can be found \cite{Ivanovic,Wootters 2}. If the Hilbert space
dimension is the product of at least two different prime numbers, it
is still unknown how many MUBs exist in the space, although there
are some approaches to try to find the solution to this problem in
some special cases, such as $\Dim=6$ or when $\Dim$ is a nonprime
integer squared \cite{Grassl,Archer,Wocjan}.

If we make a measurement of observable $\M$ on the system,
characterized by the state vector $\ketstate$, we get the
projection probabilities \beq \Pm \equiv | \langle \M,m \ketstate
|^2, \ m=1, \ldots , \Dim . \eeq From these, following
\cite{Larsen,Luis}, we define the degree of certainty $\CM$ with
which we can estimate the observable $\M$ from \beq \CMtwo \equiv
\sum_{m=1}^\Dim \Pmtwo . \eeq From the definition follows that $1
\geq \CMtwo \geq 1/\Dim$, where the two bounds are saturated by
the eigenstates of observable $\M$ and, e.g., the mutually
unbiased observable $\K$, respectively. The reason we use
certainty, rather than uncertainty, to quantify our ability to
infer some observable for a given state preparation is that the
corresponding complementarity relations take a simple form. The
following three relations were proven in \cite{Larsen} and
\cite{Luis}:

(I) For any two mutually unbiased observables $\K$ and $\M$\beq
\CMtwo + \CKtwo \leq 1 + \frac{1}{\Dim} . \label{Eq:pair
inequality}\eeq (II) For any number $J$ of mutually unbiased
observables we have \beq \Ctwo{1} + \ldots + \Ctwo{J} \leq 1 +
\frac{J-1}{\sqrt{\Dim}} . \label{Eq:J inequality} \eeq (III) If
the Hilbert-space dimension $\levels$ for each particle is a
prime, the sharper inequality \beq \Ctwo{1} + \ldots + \Ctwo{J}
\leq 1 + \frac{J-1}{\Dim} \label{Eq:prime inequality}\eeq is
valid. In this case, the maximum number of MUB are $\Dim + 1$, so
for the full set of MUB the following inequality is valid: \beq
\Ctwo{1} + \ldots + \Ctwo{\Dim + 1} \leq 2 . \label{Eq:MUB
inequality}\eeq All the equalities convey the same basic message,
the higher degree of certainty one obtains of the outcome of one
observable, the larger the uncertainty of the outcome of every
other observable becomes. This is both true for any pair of
mutually unbiased observables, Eq. (\ref{Eq:pair inequality}), and
for any set of such observables, Eqs. (\ref{Eq:J
inequality})-(\ref{Eq:MUB inequality}). Before ending this section
it is worth pointing out that there also exist bounds for the
products of certainties of MUB that may be useful
\cite{Larsen,Luis}, but we shall ignore these in the following.

\section{Complementarity between local and nonlocal qubit observables}
\label{Sec: qubits}

The simplest system to demonstrate the ladder of complementarity
relations on are qubits. The case of two qubits has already been
investigated \cite{Wootters 2,Lawrence}, but for completeness we
recapitulate the results here. Hence, we consider the case where
$\levels = 2$, $\particles = 2$, and $\Dim=4$. In this case one
can find five mutually unbiased observables. As shown in
\cite{Lawrence}, the observables come in two categories. Three of
the observables have separable eigenvectors. That is, they
represent two local measurements, each with two possible outcomes.
Examples of such mutually unbiased observables are the spin
observables $\opx \otimes \opx$, $\opy \otimes \opy$, and $\opz
\otimes \opz$. The remaining two observables have nonseparable
eigenvectors. The corresponding measurements must be done jointly
on the two qubits, i.e., employing two different kinds of
Bell-state analyzers. These nonlocal measurements each have four
outcomes. Any unitary transformation, local or nonlocal, of all
the observables will keep this structure, with three separable and
two nonseparable observables, although the observables will
change, in general \cite{Romero}. If we consider the eigenstates
of $\opz$, both the $\opx$ and the $\opy$ observable can be seen
as measurements of interference between the two eigenstates of
$\opz$. If we interpret the $\opz$ and measuring some ``particle''
characteristics, then $\opx$ and $\opy$ can be interpreted as
measuring ``wave'' characteristics \cite{Englert 1,Bjork 2}. The
single particle observables $\opx$, $\opy$, and $\opz$ are, not
unexpectedly, mutually unbiased. In addition, as demonstrated in
\cite{Jaeger 2}, and by Eq. (\ref{Eq:pair inequality}), above,
there exists a complementary relation between the single particle
interference (probed by $\opx$ and $\opy$ if the eigenstates of
$\opz$ is taken as the ``particle'' states), and the two-particle
interference probed by an appropriate Bell-state analyzer (in such
a way that the two measurements are mutually unbiased). In fact,
as expressed by Eq. (\ref{Eq:MUB inequality}), all the five
observables, where three measure local particle or wave
characteristics, and the other two measure nonlocal interferences,
jointly obey a complementarity relation. Certainty of the outcome
of one of the observables imply total uncertainty with the respect
to the other four.

Next, consider the case of three qubits. In this case, there are
three categories of observables; fully separable, biseparable, and
nonseparable. Within these three categories, five different
factorization classes exist, because the biseparable observables
may be biseparable in three different ways. If one labels the
qubits $A$, $B$, and $C$ we can express the three partitions
$A(BC)$, $(AB)C$, and $(AC)B$. In the three-qubit space there
exists four different MUB structures (with respect to
separability). Of the four structures, each defined by $8+1=9$
observables, one exist where all the nine observables are
biseparable. Another possibility is to have one fully separable,
six biseparable, and two nonseparable observables. The third
possibility is to have two fully separable, three biseparable, and
four nonseparable observables. The fourth and last possibility is
to have three fully separable and six nonseparable observables. We
shall denote the four structures $(0,9,0)$, $(1,6,2)$, $(2,3,4)$,
and $(3,0,6)$. In the context of complementarity between local and
nonlocal observables, the $(1,6,2)$ and $(2,3,4)$ MUB structures
are the most interesting, because they contain observables
representing each separability category. In fact, both structures
contain at least one observable corresponding to each of the five
factorization classes.

Let us focus on the $(2,3,4)$ structure. Of the fully separable
observables, one, say $\opz \otimes \opz \otimes \opz$ can be said
to measure some ``particle'' characteristics of each qubit
separately. If so, the mutually unbiased observable $\opx \otimes
\opx \otimes \opx$ will project onto superpositions of the
eigenstates to the ``particle'' observable, so $\opx \otimes \opx
\otimes \opx$ can be considered to measure local ``wave''
characteristics. In this case, it is clear that the factorable
single qubit observable in each of the three biseparable
observables must be $\opy$ in order for the observables to be
mutually unbiased to the fully separable observables. Hence, the
three biseparable observables correspond to a single particle
``wave interference'' measurement combined with a Bell-state
analysis of the remaining two qubits. The last four observables
are nonseparable and correspond to the projections on four
mutually unbiased, complete sets, of GHZ-states. Again, through
Eq. (\ref{Eq:pair inequality})-(\ref{Eq:MUB inequality}), it is
clear that there exist complementary relations between any subset
of the nine observables. The sharpest of the relations is
(\ref{Eq:MUB inequality}), which tells us that an increase of the
degree of certainty of one of the observables in general must be
``paid for'' by decreasing the degree of certainty of the
remaining observables.

If we consider four qubits (defined on a 16-dimensional Hilbert
space) the complexity increases somewhat. The full set of mutually
unbiased observables contain $16+1=17$ operators. Five categories of
observables exist; fully separable, triseparable, biseparable into a
$4 \times 4$ space, biseparable into a $2 \times 8$ space, and
nonseparable. Within these five categories, 18 different
factorization classes exist, because the triseparable observables
may be factored in six different ways, the $4 \times 4$ biseparable
operators may also be factored in six different ways, and the $2
\times 8$ biseparable operators may be factored in four different
ways. In total, one can show that 16 different MUB structures exist,
but only four of them contain all the five categories of
separability just delineated \cite{Romero}. If we follow the
notation above, and label the MUB structure by denoting the number
of fully separable, triseparable, $\ldots$ , nonseparable
observables, these four structures are $(2,1,2,2,10)$,
$(1,1,6,2,7)$, $(1,2,4,2,8)$, and $(1,3,2,2,9)$. Of course, the 17
observables cannot represent every of the $18$ different
factorization classes. None the less, for any of the observables,
may it be fully separable (i.e., local to every qubit), partially
separable, or nonseparable (i.e., a joint measurement of all
qubits), full certainty of the measurement outcome will imply total
uncertainty of all of the 16 other observables. Quite obviously, two
observables belonging to different factorization classes cannot
commute, irrespective if they are mutually unbiased or not.
Therefore, probing different kinds of non-local properties is bound
to be limited by complementarity to some extent. The MUB sets of
observables are the ones that take the complementarity to the limit,
as manifested by Eq. (\ref{Eq:MUB inequality}). In this sense the
observables correspond to the optimal, discrete, state-defining
propositions, where most of them probe non-local properties, as
suggested by Zeilinger \cite{Zeilinger}.

\section{Complementarity between local and nonlocal qutrit observables}

Two qutrits represent the case where $\levels=3$, $\particles=2$ and
$\Dim=9$. It has been shown by Lawrence, that in this Hilbert space
only one MUB structure exist, and it has four separable and six
nonseparable observables \cite{Lawrence 2}. This case is therefore
similar to the two qubit case. Hence, local and non-local
observables do not commute, and the MUB set of observables,
containing both kinds of observables, is the ``maximally
noncommuting'' set.

If we increase the number of qutrits to three, we have already
seen in Sec. \ref{Sec: qubits} that three separability categories
and five factorization classes exist. One can deduce from
\cite{Lawrence 2} that there exist five MUB structures; (0,12,16),
(1,9,18), (2,6,20), (3,3,22), and (4,0,24). Of these, the
(1,9,18), (2,6,20), and (3,3,22) structures contain observables
representing all five factorization classes. Hence, all
conceivable types of local and non-local properties are covered by
the certainty relation (\ref{Eq:MUB inequality}).

\section{The general case}

So far, we have looked at subsystems (or particles) defined on a
Hilbert space of prime dimension, where particularly restrictive
certainty relations exist. What happens if $\levels$ is composite?
The answer is that, in fact, little changes. For the sake of
concreteness, let us look at the case of three subsystems, each
defined on a six-dimensional Hilbert space. The composite system
space dimension can be written $\Dim = 6^3 = (2 \times 3)^3 = 2^3
\times 3^3$. Again, three separability categories and five
factorization classes exist. We can now take the one of the two MUB
structures (defined by nine observables) containing all
factorization classes for three qubits, and tensor multiply it by a
selection of nine of the 28 MUB observables that represent similar
factorization classes for three qutrits. In this way we obtain nine
MUB observables defined on the 216-dimensional composite space, and
this set will cover all the five different factorization classes of
the three particles. We see that subsystems defined on a
composite-dimensional space will inherit most of the desirable (and
also the possibly undesirable) properties from the most restrictive
subspace $\levels$ can be factored into. The main difference between
prime- and nonprime-dimensional particles is that we can no longer
use the more restrictive certainty relation (\ref{Eq:MUB
inequality}), but are left with the less restrictive inequality
(\ref{Eq:J inequality}).

\section{Conclusions}

We have demonstrated that for any number of particles, each
defined on a Hilbert space of arbitrary dimension, a whole set of
complementarity relations between local and nonlocal properties
exist. Between any factorization of the particles, corresponding
to joint measurements of certain groups of particles and
individual measurements of other particles, one can find pairs and
whole sets of observables that obey rather tight complementarity
relations. It is, of course, possible to find other pairs, or
sets, of observables corresponding to any particular factorization
of Hilbert space that are not as strongly confined by a certainty
or uncertainty relation as the MUB observables are.

We have used certainty as the information measure of each
observable, as these complementarity relations take on a simple
and \ae sthetically pleasing  form. Other forms of complementarity
relations between the same, or similar, observables can probably
also be derived, although we have made no such attempt in this
work.

\acknowledgments{This work was supported by the Swedish Institute
(SI), the Swedish Foundation for International Cooperation in
Research and Higher Education (STINT), the Swedish Foundation for
Strategic Research (SSF), and the Swedish Research Council (VR).}

\end{document}